# Impostor Phenomenon in Software Engineers

Paloma Guenes
Pontifical Catholic University of Rio de Janeiro (PUC-Rio)
Rio de Janeiro, Brazil
palomaguenes@gmail.com

Rafael Tomaz
Pontifical Catholic University of Rio de Janeiro (PUC-Rio)
Rio de Janeiro, Brazil
rafaelstomaz@gmail.com

Marcos Kalinowski
Pontifical Catholic University of Rio de Janeiro (PUC-Rio)
Rio de Janeiro, Brazil
kalinowski@inf.puc-rio.br

Maria Teresa Baldassarre
University of Bari
Bari, Italy
mariateresa.baldassarre@uniba.it

Margaret-Anne Storey
University of Victoria
Victoria, Canada
mstorey@uvic.ca

## ABSTRACT

The Impostor Phenomenon (IP) is widely discussed in Science, Technology, Engineering, and Mathematics (STEM) and has been evaluated in Computer Science students. However, formal research on IP in software engineers has yet to be conducted, although its impacts may lead to mental disorders such as depression and burnout. This study describes a survey that investigates the extent of impostor feelings in software engineers, considering aspects such as gender, race/ethnicity, and roles. Furthermore, we investigate the influence of IP on their perceived productivity. The survey instrument was designed using a theory-driven approach and included demographic questions, an internationally validated IP scale, and questions for measuring perceived productivity based on the SPACE framework constructs. The survey was sent to companies operating in various business sectors. Data analysis used bootstrapping with resampling to calculate confidence intervals and Mann-Whitney statistical significance testing for assessing the hypotheses. We received responses from 624 software engineers from 26 countries. The bootstrapping results reveal that a proportion of 52.7% of software engineers experience frequent to intense levels of IP and that women suffer at a significantly higher proportion (60.6%) than men (48.8%). Regarding race/ethnicity, we observed more frequent impostor feelings in Asian (67.9%) and Black (65.1%) than in White (50.0%) software engineers. We also observed that the presence of IP is less common among individuals who are married and have children. Moreover, the prevalence of IP showed a statistically significant negative effect on the perceived productivity for all SPACE framework constructs. The evidence relating IP to software engineers provides a starting point to help organizations find ways to raise awareness of the problem and improve the emotional skills of software professionals.

## 1 INTRODUCTION

Have you ever found yourself having doubts about your abilities or expertise when others around you indicate otherwise? Have you ever denied praise for something you think was too easy or where you believe that anyone else could have done a better job? Recent studies reveal that many software engineers often experience negative feelings such as frustration and unhappiness when they think about their own perception of their abilities or skills [8, 11, 13].

In 1978, Pauline Clance and Suzanne Imes [6] identified the Impostor Phenomenon (IP)[1] in high-achieving female individuals. IP corresponds to the feeling of not recognizing oneself accurately and frequently facing a great fear of being discovered as a fraud. Later, Dr. Clance created a scale (the Clance IP Scale - CIPS) [5] to determine to which extent individuals suffer from IP.

Based on the current literature, we know that these feelings are shared among Computer and Data Science students [7, 22, 26]. The commonality among these studies is that over half of the students suffer from IP. In spite of the evidence presented by these three studies, there is no scientific confirmation that impostor feelings persist in the professional life of people with a degree in Computer Science (CS). Moreover, Dr. Clance in 1978 had already observed that impostor feelings predominantly afflict students among the general population. Therefore, there is no clear evidence that IP also manifests in software engineers.

---

[1]The term *Impostor Syndrome* is also used to refer to Dr. Clance's work. However, the terminology "syndrome" is related to an official medical diagnosis, while IP is not. In this paper, we acknowledge the official name Impostor Phenomenon while also considering research that uses the term *Impostor Syndrome*.

Gathering user requirements and transforming them into software is the responsibility of software engineers. Their activities may relate to any phase of a software development process [14], which, being challenging, could lead to feelings of being an impostor. They typically know they must be ready to constantly overcome challenges intrinsic to the profession while coding new features or fixing bugs. Whereas succeeding in a task is a moment of joy experienced by developers, getting stuck and the dread of failure and frustration are prejudicial to productivity [13]. The consequences of IP [4, 6] perceived by software engineers include anxiety, burnout, and depression [12].

This paper aims to investigate the presence of IP in software engineers and its impact on their perceived productivity through a survey. The survey instrument was designed using a theory-driven approach. We identified two main research questions: *RQ1 - How does IP manifest in software engineers?* and *RQ2 - Does IP affect the perceived productivity of software engineers?* We detail these research questions and our derived hypotheses to be assessed. Our survey instrument was built to support the testing of these hypotheses. For measuring the IP, we obtained authorization from Dr. Clance to use the internationally validated CIPS scale [5]. For the perceived productivity, similar to other researchers [3], we derived statements to measure the constructs of the SPACE developer productivity framework (Satisfaction and well-being, Performance, Activity, Communication and collaboration, and Efficiency and flow) [9, 10] using a five-point Likert-scale.

We were granted authorization to use the CIPS scale within our survey using a closed invitation format, i.e., directly addressing the target population. We reached out to over 100 companies worldwide of different sizes and different business domains, asking them to distribute the survey internally to their software engineers. We received responses from 624 software engineers from 26 countries, a sample size that allows us to achieve conclusion validity. Furthermore, to strengthen our confidence in the representativeness of our sample, we compared it against the sample of the Stack Overflow Annual Developer survey of 2023 [20] and observed comparable distributions for the main variables we considered.

Our main findings indicate that 52.72% of software engineers suffer from frequent to intense levels of IP. Analyzing underrepresented groups sheds light on alarming differences. For instance, women suffer from IP at a significantly higher rate (60,64%) than men (48,82%). Furthermore, Asian (67.85%) and Black (65.11%) suffer more than White (50.00%) software engineers. We also observed that the presence of IP is less common in individuals who are married and have children. With respect to perceived productivity, the prevalence of IP showed a statistically significant negative effect on all five SPACE developer productivity framework constructs (as presented later).

The rest of this paper is organized as follows. Section 2 provides the background on the Impostor Phenomenon, related work, and introduces the SPACE developer productivity framework. Section 3 details our goal and research questions. Section 4 explains our methods for designing the survey, data collection, and data analysis. Section 5 presents the results. Section 6 discusses the results and Section 7 the threats to validity of our study. Finally, Section 8 concludes the paper.

## 2 BACKGROUND AND RELATED WORK

In this section, we provide the theoretical background for our research and review related work. First, we describe the IP and psychometric instruments for its measurement. Thereafter, as the literature still does not characterize IP in software engineers, we report on studies investigating IP in computer science and data science students (the closest research to our topic of investigation). Finally, we briefly introduce the SPACE of Developer Productivity framework, which we use to investigate the potential impacts of IP on perceived productivity and well-being.

### 2.1 Impostor Phenomenon

As originally defined, the Impostor Phenomenon is the experience of intellectual phoniness perceived by high-achieving professionals [6]. These individuals have a great fear that others will discover that they are not as competent as they appear, attributing their successes to luck, meeting the right people, being in the right place at the right time, or even their personal charm [6].

In 1985, Dr. Clance, one of the psychologists who first identified this phenomenon in high-achieving women, created a scale to determine if a person is suffering from IP and to what extent [5]. While there are four scales to determine IP (Clance Impostor Phenomenon Scale, Harvey Impostor Scale, Perceived Fraudulence Scale, and Leary Impostor Scale), Clance Impostor Phenomenon Scale (CIPS) is the most used scale by researchers and practitioners [18].

The CIPS scale consists of a 20-item questionnaire. Each item concerns a statement that is rated on a Likert scale from 1 to 5. A rating of 1 corresponds to "Not at all true" or total disagreement with the aforementioned statement, and a rating of 5 corresponds to "Very true" or complete agreement.

Upon completing the questionnaire, it is necessary to sum the values of each response in order to obtain the scale value results. The higher the scale value, the more often and seriously IP interferes with a person's life. According to the CIPS scale, scoring more than 60 points represents that the respondent has frequent and intense IP feelings, which means meeting the diagnostic criteria used in previous research with students (discussed in Section 2.2).

### 2.2 Impostor Phenomenon in Computer and Data Science Students

There are two recent studies that investigate IP in Computer Science students, both using the CIPS scale and these were conducted in research-intensive environments.

In 2020, Rosenstein *et al.* [22] conducted a study in a North American institution and found that among over 200 students, 57% suffered from frequent and intense IP. Therefore, respondents experience frequent to intense IP feelings. Besides, the study concludes that IP in Computer Science students happens more often than in other groups from comparable studies. They also observed that IP was particularly prevalent in women (of which 71% suffered from frequent and intense levels of IP).

Two years later, Zavaleta *et al.* [26] in a newer study on Computer Science students corroborated the previous one (mentioned above). They found a larger number of students suffering from IP (68%) and, again, women reported higher CIPS scores than men (78%). The diagnostic criterion in this study was scoring above 60.

Recently, a study conducted with 86 master students of data science [7] from three different universities found out that 53% of students have intense or frequent feelings of IP. However, they have not found significant differences in IP related to gender or race.

Even though there might be a natural connection between computer science, data science students, and software engineers, the professional context is different and there is currently no study investigating the prevalence and manifestation of IP in software engineers.

In this study, we also consider the diagnostic criterion as scoring above 60, which means having frequent to intense levels of impostor feelings.

## 2.3 SPACE of Developer Productivity

Forsgren *et al.* recently created the SPACE of Developer Productivity framework [9, 10]. This framework considers five dimensions of productivity, including the individual's perspective and how the developer deals with the work environment: the amount and quality of tasks developed and teamwork, among others. This framework is widely used in industry for measuring productivity and well-being, for example see the research on the impact of AI on developer productivity [3]. A short description of the SPACE dimensions follows, based on [10].

- Satisfaction and well-being: Satisfaction is how fulfilled developers feel with their work, team, tools, or culture; well-being is how healthy and happy they are and how their work impacts it.
- Performance: Refers to the outcome of a system or process. It is related to quality, reliability, absence of bugs, ongoing service health and impact, customer satisfaction, customer adoption and retention, feature usage, and cost reduction.
- Activity: Refers to the count of actions or outputs (e.g., issues, code reviews) completed in the course of performing work.
- Communication and collaboration: Captures how people and teams communicate and work together.
- Efficiency and flow: Capture the ability for an engineer to complete their work or make progress on it with minimal interruptions or delays, whether individually or through a system.

Productivity and performance are the areas most affected by the dissatisfaction of developers [13]. In this paper, we posit that individuals who view themselves as impostors also tend to perceive their productivity as lower than their peers. We propose to use the SPACE framework to understand the perceived productivity of software engineers and its relation with IP.

## 3 GOAL AND RESEARCH QUESTIONS

Our research goal can be stated according to the Goal-Question-Metric paradigm goal definition template [2] as follows; *Analyze* the prevalence of Impostor Phenomenon in software engineers *with the purpose of* characterizing *with respect to* the prevalence of the phenomenon in different roles and profiles (e.g., gender, race/ethnicity) and its relation with perceived productivity *from the point of view* of the researcher *in the context of* software engineering professionals.

From this goal, we derived and detailed two Research Questions (RQs) addressed through a survey instrument:

- RQ1: How does IP manifest in software engineers?
  - RQ1.1: What proportion of Software Engineers suffer from IP?
  - RQ1.2: How does IP manifest in different genders, races/ethnicities, and roles?
- RQ2: Does IP affect the perceived productivity of software engineers?
  - RQ2.1: Does IP affect Satisfaction and Well-being?
  - RQ2.2: Does IP affect Performance?
  - RQ2.3: Does IP affect Activity?
  - RQ2.4: Does IP affect Communication and Collaboration?
  - RQ2.5: Does IP affect Efficiency and Flow?

In order to answer RQ1.1, we used the CIPS scale. For RQ 1.2, we applied the blocking principle to the results on the IP prevalence based on the demographic questions. To answer RQ2, we created a group of questions to understand the developer's perceived productivity, considering the five dimensions of the SPACE framework. Further details on the survey design follow.

## 4 SURVEY DESIGN

We used a theory-driven survey design approach [24], in which we first hypothesize on the theory to be assessed and then elaborate the survey instrument by selecting validated scales for the different theoretical constructs. Our first hypothesis, related to RQ1, concerns the overall prevalence of IP in software engineers. Hypotheses 2 to 6 were formulated to investigate RQ2.

*Hypothesis H1: More than 50% of software engineers suffer from the Impostor Phenomenon*. More than half of computer and data science students were tested and identified as suffering from frequent to intense levels of IP meeting the diagnostic criteria [22, 26]. It seems reasonable to assume that the fear of being discovered as a fraud will not disappear once they graduate and start a professional life. Also, we want to understand the prevalence of IP within specific groups of software engineering professionals, in particular, per gender, race/ethnicity, and role.

Furthermore, we want to assess if there is a difference for software engineers suffering from IP on their perceived productivity using the SPACE developer productivity framework dimensions as constructs. We assume that people who consider themselves a fraud also perceive themselves as less productive than others. We elaborate on Hypotheses 2 to 6 based on this informal deductive intuition and Dr. Clance's book [5].

*Hypothesis H2: Software engineers that meet the diagnostic criterion of IP have lower perceived satisfaction and well-being.* The second hypothesis is related to the satisfaction and well-being productivity construct. Forsgren *et al.* [9] declare that low productivity is related to low personal satisfaction. We hypothesize that there is a relationship between suffering from IP and perceived personal satisfaction and well-being.

*Hypothesis H3: Software engineers that meet the diagnostic criterion of IP have lower perceived performance*. We derived H3 based on the Impostor Cycle defined by Dr. Clance [5], which states that the feeling of gratefully succeeding in a task is quickly transposed by the feeling that if they could do it, anyone else could perform the same way or even better. Therefore, people suffering from IP may not perceive that they deliver high-quality and impactful work even if they receive acknowledgment.

*Hypothesis H4: Software engineers that meet the diagnostic criterion of IP have lower perceived activity.* We theorize that IP feelings prevent people from solving a number of work items, pull requests, code reviews, etc., because they tend to remember the difficult times more vividly than their accomplishments in tasks completed with ease. They often focus on unfamiliar details rather than promptly applying their existing knowledge [5].

*Hypothesis H5: Software engineers that meet the diagnostic criterion of IP have lower perceived communication and collaboration.* We believe that communication might be impaired because of fear of judgment, being exposed as fraud, and constantly worrying about not matching others' expectations. That is, since they tend to believe more in others than in themselves, they tend not to give their opinion even about subjects where they arespecialists [5].

*Hypothesis H6: Software engineers that meet the diagnostic criterion of IP have lower efficiency and flow.* We hypothesize that efficiency and flow are rarely perceived by individuals with IP. In general, feelings of being stuck or held back are common in individuals with IP because they tend to doubt themselves constantly [5].

### 4.1  Instrument Design

To design the survey instrument, first, we included demographic and filter questions of our interest to improve criterion validity and to allow us to investigate how the IP manifests in different groups, including underrepresented ones. These questions concern objectively gathering information on the gender, race/ethnicity, role, age, level of education, and experience.

Additionally, a study focused on Open Source Software shows that women in this context have half as many kids as men [21], and very few are married or cohabiting with partners [19]. To understand if IP is related to these situations, we added two more questions asking about marital status and the number of children.

For the substantive questions, we considered the CIPS scale as the psychometric instrument for assessing IP [5]. The main reason for this decision was that it is a widely accepted and validated scale and the most used one by researchers and practitioners [18]. Furthermore, this scale allows the comparison of our results with the previous studies conducted with computer and data science students (*cf.* Section 2.2).

The CIPS score is divided into four scoring categories: 40 or less, representing few impostor characteristics; 41 to 60, representing moderate IP experiences; 61 to 80, representing frequently having impostor feelings; and 80 or more, representing intense IP experiences. Scoring more than 60 means meeting the diagnostic criterion.

For the productivity constructs involved in H2 to H6, we used the definition of these constructs provided by the SPACE framework [9] and phrased Likert-scale self-assessment statements, ranging from 1 (strongly disagree) to 5 (strongly agree). Table 1 details the 10 created questions. We reviewed the questions with experts on the SPACE framework. At the very end of the survey, we added an open-ended question allowing the participants to share any additional comment or experience.

Regarding ethics, we followed recommended procedures to obtain consent for empirical studies in software engineering [1]. We embedded an informed consent form in our survey, communicating the purpose of the research, importance of the research and rigor, procedures, voluntariness, right to terminate, benefits/risks, and assuring anonymity and confidentiality. The research plan, the informed consent form, and the survey were submitted to the university ethics board for approval and adjusted until they entirely met the required criteria.

We implemented the survey in a tool called Tally, which is secure for data transmission and is GDPR compliant[2]. The authorization to use the CIPS scale, the consent form we designed, the complete survey instrument we used, our anonymized raw data, and our analysis scripts can be found in our online open science repository[3].

### 4.2  Data Collection

As mentioned, we formally obtained the required authorization from Dr. Clance to use the CIPS scale. However, the use of the scale requires surveys to have a closed invitation format, directly addressing the target population. Hence, we could not spread the survey on the open internet or social media platforms. Therefore, our method was to approach partner companies by sending e-mails to software project managers and developers with a link to access the survey and asking them to distribute it within their companies. All respondents were asked to agree with the consent form before getting access to the survey. We conducted a pilot in April 2023 with seven software engineers who have more than 3 years of work experience, representing distinct genders, ages, education levels, and ethnic backgrounds. After minor adjustments, data collection happened from May 2023 to July 2023.

### 4.3  Data Analysis Procedures

To discuss the representativeness of our sample, we compare the characteristics of our sample (e.g., age range, company size, and working experience) with data from Stack Overflow's software developer survey [20]. Annually, they conduct a survey among their users to gain insights into the software engineering community. This year, their survey received more than 90,000 answers[20].

To assess the manifestation of the IP in software engineers, we calculate confidence intervals using a technique called Bootstrapping that has been reported to be more reliable and precise than statistical inferences drawn directly from samples [16, 24]. Bootstrapping calculates confidence intervals by re-sampling our data set, creating many simulated samples to promote a more robust and accurate analysis. Considering our sample size N, to perform bootstrapping, we create new samples S times of the same size N. Re-samples may include a given response zero or more times. We set S to 1000, a value known to yield meaningful statistical results [16]. We also calculate simple frequencies for the prevalence of IP in different races/ethnicities, roles, educational levels, years of experience, marital status, and number of children. We created scripts in Python to support the analysis of our sample. The CIPS scale provides a score representing the intensity of IP for each respondent. We created a Boolean column to explicitly indicate whether the person had met the IP criterion aiming to streamline the subsequent analysis.

---

[2] https://tally.so/help/gdpr
[3] https://doi.org/10.5281/zenodo.8415205

Table 1: SPACE based questions

| Research Question | SPACE Dimension | Survey Question |
|---|---|---|
| RQ2.1 | Satisfaction and Well-Being | 1. I feel fulfilled at my work. <br> 2. I am healthy and happy when I work. |
| RQ2.2 | Performance | 3. I deliver high quality work. <br> 4. I deliver impactful work. |
| RQ2.3 | Activity | 5. I complete as many tasks as expected from me in my position. |
| RQ2.4 | Communication and Collaboration | 6. I feel comfortable communicating with my team. <br> 7. My team supports and values my communication. <br> 8. I play an important role in my team. |
| RQ2.5 | Efficiency and Flow | 9. I am able to focus on and make progress on my work without internal interruptions (eg., mind wandering, lack of confidence). <br> 10. I am able to focus on and make progress on my work without external interruptions (eg., notification from mobile device, a colleague asking a question). |

Finally, with respect to the relation of IP with perceived productivity, we applied the nonparametric Mann-Whitney U-test to check for statistically significant differences (alpha value 0.05) in the answers to the statements of Table 1. This test can be safely applied when one variable is nominal (e.g., having met the IP criterion or not) and one is ordinal (e.g., our Likert-scale questions). The anonymized raw data and our Python analysis scripts can be found in our online open science repository[3].

## 5  SURVEY RESULTS
### 5.1  Study Population

Similar to Wagner *et al.* [24] and according to Yamane's equation to calculate a suitable sample size [25], considering the worldwide developer population to be 26.3M developers[4], a sample size of N=400 software engineers would be sufficient to allow generalizability for most purposes, a criterion that we achieved successfully with 624 answers. By distributing the survey to key contacts from over 100 different companies, we were able to reach 26 countries from five continents, as shown in Figure 1. The 16 *Others* represent two participants from Cuba, Finland, and Romania and one participant from Argentina, Austria, China, Cyprus, Czech Republic, Pakistan, Peru, Russian Federation, Senegal, and Slovakia. All survey questions were mandatory except the open-ended question for which 67 respondents left a comment. One-third of our respondents work in companies exceeding ten thousand employees.

---
[4]Updated numbers from Evans Data Corporation. Site: www.evansdata.com

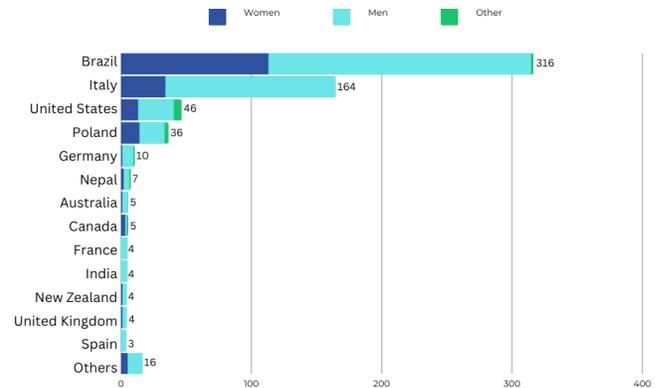

Figure 1: Participants' countries.

The results of the demographic questions are shown in Table 2. Men represent 67.63% of our sample, while women constitute 30.13%. The most common ethnicity of software engineers in our sample is *White* (78.21%), followed by *Black or African American* (6.89%) and *Asian* (4,49%). *Other* represents *American Indian or Alaska Native*, *Other Race*, and *Prefer not to answer*. With respect to the level of education, the majority (58.33%) hold a Bachelor's degree or equivalent. The sample also shows a balance between married (49.68%) and single (44.23%) respondents. Finally, the majority (68.75%) of our sample reported not having children.

Figure 2 shows a word cloud representation of the most common business sectors. *Banking/Financial* and *Sales/E-commerce* business sectors together represent 30% of our sample. We received a significant number (202) of responses in the *Other* text field, primarily encompassing *Consulting* and the *Public Sector*.

**Table 2: Demographic questions.**

| | | |
|---|---|---|
| What is your gender? | Male | 67.63% |
| | Female | 30.13% |
| | Other | 1.44% |
| | Prefer not to answer | 0.8% |
| What is your predominant race/ethnicity? | White | 78.21% |
| | Black or African American | 6.89% |
| | Asian | 4.49% |
| | Other | 10.42% |
| What is your level of education? | Bachelor's or equivalent level | 58.33% |
| | Master's or equivalent level | 27.40% |
| | Secondary education | 7.69% |
| | Doctoral or equivalent level | 6.41% |
| | Primary education | 0.16% |
| What is your marital status? | Married or Cohabiting | 49.68% |
| | Single | 44.23% |
| | Prefer not to answer | 3.85% |
| | Divorced | 2.08% |
| | Widow/Widower | 0.16% |
| How many children do you have? | 0 | 68.75% |
| | 1 | 14.74% |
| | 2 | 12.02% |
| | 3 or more | 2.72% |
| | Prefer not to answer | 1.76% |

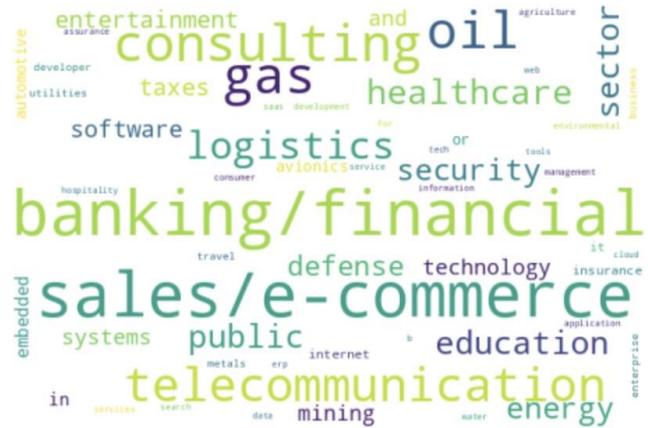

**Figure 2: Business sectors.**

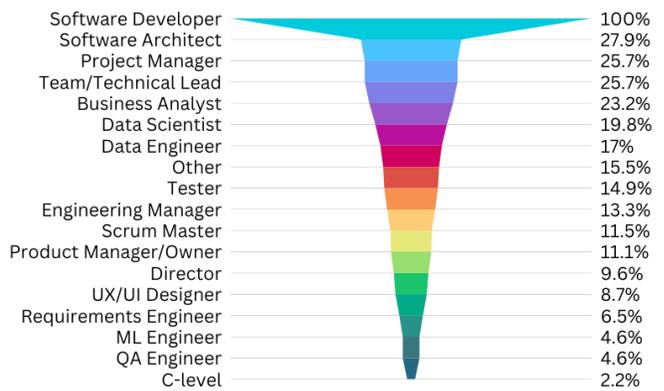

**Figure 3: Participants' roles.**

Regarding the roles of the respondents, software developers (323) make up the largest portion of our sample, as shown in Figure 3. It is noteworthy that respondents could report to be working in more than one role. The *Other* role was described in a text field by 39 respondents. Consultant, Cyber Security Engineer, and Analyst were the most common within these answers.

For the purpose of assessing the representativeness of our sample, we conducted a comparative analysis with data from the Stack Overflow Annual Developer Survey 2023 [20]. Figure 4 portrays the distribution of age range, years of working experience, and company size from both studies. Comparisons shown in Figure 4a were facilitated by using similar age categories. To calculate the distributions shown in Figures 4b and 4c, we used the raw data available from Stack Overflow [20].

It is possible to observe that the distributions of the study were similar considering age range (see Figure 4a), working experience, and company size addressed in both studies. There are a few differences. For instance, in Fig. 4b, we observe that our sample had a greater representation of participants in the early stages of their careers. Also, Figure 4c allows us to observe that our sample has a higher representation of individuals from companies with over 10k employees and a lower representation from companies with less than 9 employees. It is noteworthy that the Stack Overflow survey in its question *Approximately how many people are employed by the company or organization you currently work for?* had a specific option for freelancers, while our restricted data collection, required for using the CIPS scale, made it difficult to reach out to professionals that were not working for a specific company. For the same reason, our geographic distribution was naturally different, as we had to personally reach out to companies we had contact with. Overall, the similarities of the samples exceed our expectations and increase the confidence in the sample's representativeness, particularly considering that the Stack Overflow survey sample comprises over 90k responses.

Another difference is that 30.13% of our answers were provided by women, significantly more than in the 2023 Stack Overflow survey (5.17%). Greater female participation in our sample can possibly be attributed to the fact that we do not consider only developers (who are the main users of Stack Overflow). We consider software engineering professionals involved in different roles in the software process.

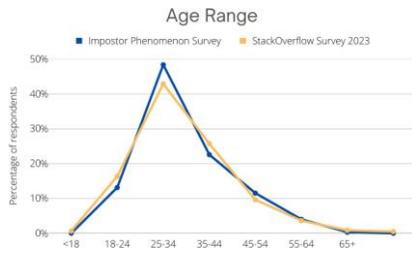

(a) Age range distribution from this study and the 2023 Stack Overflow survey.

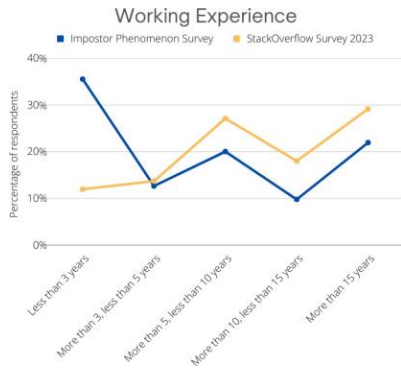

(b) Working experience distribution from this study and the 2023 Stack Overflow survey.

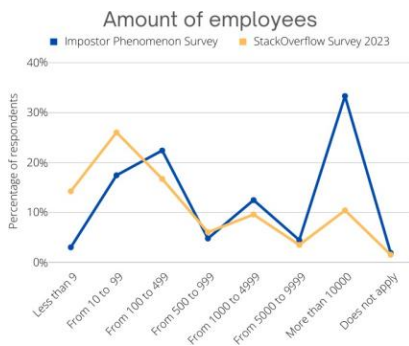

(c) Company size distribution from this study and the 2023 Stack Overflow survey.

Figure 4: Sample comparison of this study and the Stack Overflow Annual Developer Survey 2023.

## 5.2 RQ1 - How does IP manifest in software engineers?

Answering RQ1.1, the proportion of software engineers who suffer from IP is shown in Figure 5, together with an error bar that represents the 95% confidence interval. Both the Proportion (P) and the Confidence Intervals (CI) were calculated using bootstrapping [24].

The results show that P = 52.72% (CI [48.72, 56.57]) match the diagnostic criteria of IP, suffering from frequent to intense levels of impostor feelings. Therefore, while we observe a high proportion of software engineers suffering from IP, considering the confidence interval, we cannot confirm hypothesis H1, which states that more than 50% of software engineers suffer from IP. However, software engineers' mean IP score is 62.12, which is high and comparable to the mean score of 64.18 observed among Computer Science students [22].

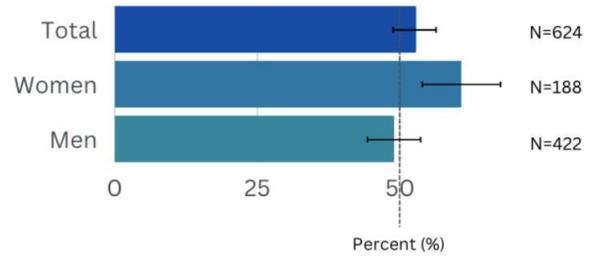

Figure 5: Bootstrapping proportions of IP with confidence intervals.

RQ1.2 concerns the manifestation of IP within different genders, races/ethnicities, and roles. With respect to genders, Figure 5 shows that significantly higher proportions of women (P = 60.64% [53.72, 67.55]) suffer from IP than men (P = 48.82% [44.06, 53.55]). This allows us to confirm H1 for women, but not for software engineers in general nor for men. Unfortunately, our sample includes a very limited representation of genders other than male or female, not allowing for statistical bootstrapping. Therefore, we only report the simple frequency. From 14 answers (including *Prefer not to answer*), nine met the diagnostic criterion, meaning a (high) frequency of 64.28% within these gender groups.

Regarding the prevalence of IP among software engineers of different races/ethnicities, given the number of categories, we also conservatively did not apply statistical bootstrapping and limited our analysis to frequency counting without inferential statistics. The frequencies are shown in Table 3. It is noteworthy that respondents who identify predominantly as *Asian* (67.85%) and *Black or African American* (65.11%) software engineers suffer more than respondents who identify as *White* (50.00%). Hence, for races/ethnicities, we also observed differences that deserve attention from the community.

Table 3: Races/ethnicities and IP frequencies.

| Predominant Race/Ethnicity | N | IP Frequency |
|---|---|---|
| White | 488 | 50.00% |
| Prefer not to answer | 60 | 58.33% |
| Black or African American | 43 | 65.11% |
| Asian | 28 | 67.85% |
| Other Race | 3 | 100.00% |
| American Indian or Alaska Native | 2 | 0.00% |

To address how IP manifests in different roles, Figure 6 depicts the frequency of respondents matching the IP diagnostic criterion within each role. While it is not possible to draw any conclusions, it seems that technical roles, in particular, roles related to data science (*e.g.*, data scientist and data engineer), have slightly higher IP manifestation frequencies.

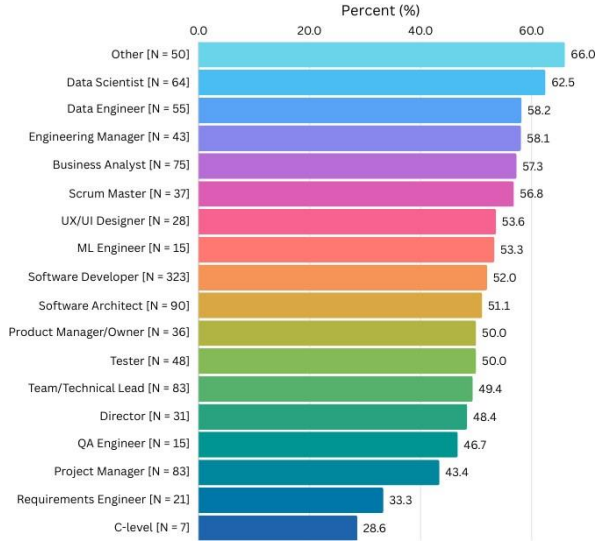

**Figure 6: Manifestation frequency of IP in different roles.**

Finally, we analyzed the IP frequencies per marital status (Table 4) and number of children (Table 5). Again, we refrain from using inferential statistics for smaller groups within the categories. Still, the frequencies lead to interesting conjectures. For instance, we observed a less common presence of IP in respondents who are married and who have one or two children. According to our data[3], this holds for men and women.

**Table 4: Marital statuses and IP frequencies.**

| Marital Status | N | Percentage with IP |
|---|---|---|
| Single | 276 | 59.78% |
| Married or Cohabiting | 310 | 47.09% |
| Prefer not to answer | 24 | 41.66% |
| Divorced | 13 | 53.84% |
| Widow/Widower | 1 | 100.00% |

**Table 5: Number of children and IP frequencies.**

| Number of Children | N | Percentage with IP |
|---|---|---|
| 0 | 429 | 57.10% |
| 1 | 92 | 45.65% |
| 2 | 75 | 33.33% |
| 3 or more | 17 | 64.70% |
| Prefer not to answer | 11 | 54.54% |

## 5.3 RQ2 - Does IP affect the perceived productivity of software engineers?

To answer RQ2, we analyze if satisfying the IP diagnostic criterion has an effect on the Likert-scale self-assessment statements (*cf.* Table 1). We applied the Mann-Whitney U-test to check for statistically significant differences (alpha value 0.05) between software engineers suffering from IP and those not suffering from it in their productivity self-assessments.

Figures 7 to 11 provide diverging stacked bar charts that enable visualizing the differences in the Likert-scale results for each statement. It is easy to observe differences with higher agreement frequencies in all statements for software engineers who do not suffer from IP. In fact, all the differences were statistically significant with extremely low p-values, and hypotheses H2 to H6 were confirmed. Software engineers suffering from frequent to intense levels of IP have lower perceived productivity.

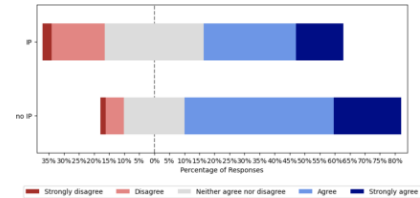

(a) **1. I feel fulfilled at my work (p-value = 3.56e-09).**

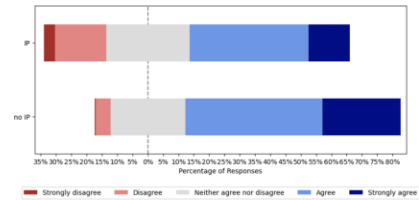

(b) **2. I am healthy and happy when I work (p-value = 3.34e-12).**

**Figure 7: Perceived satisfaction and well-being.**

## 6 DISCUSSION

The results obtained from our study offer valuable insights into the prevalence of IP among software engineers, shedding light on various aspects of this phenomenon. Hereafter, we discuss the main findings related to our research questions.

**RQ1 - How does IP manifest in software engineers?** This question explored how the IP manifests across different software engineering demographic and professional dimensions. The investigation into the overall prevalence of impostor feelings among software engineers (RQ1.1) revealed that a proportion of about 52.72% (CI [48.72, 56.57]) experience frequent to intense levels of impostor feelings. Given that IP may lead to mental disorders such as depression and burnout, this scenario highlights the need for further investigation into this phenomenon.

With regard to gender (RQ1.2), the results indicated significant differences in the manifestation of impostor feelings. A notably

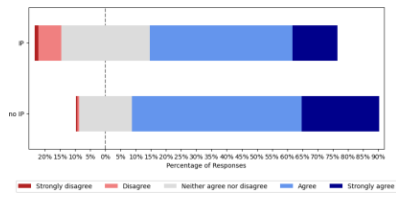

(a) 3. I deliver high-quality work (p-value = 8.78e-09).

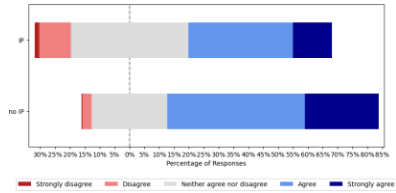

(b) 4. I deliver impactful work (p-value = 4.95e-10).

Figure 8: Perceived performance.

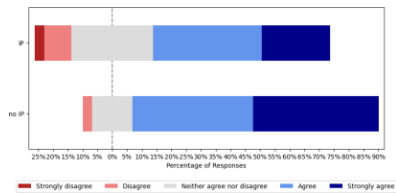

(a) 5. I complete as many tasks as it is expected from me in my position (p-value = 4.95e-10).

Figure 9: Perceived activity.

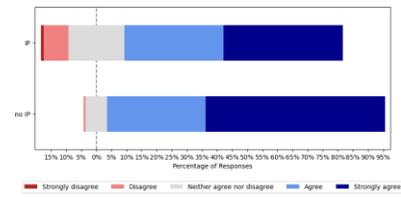

(a) 6. I feel comfortable communicating with my team (p-value = 2.38e-10).

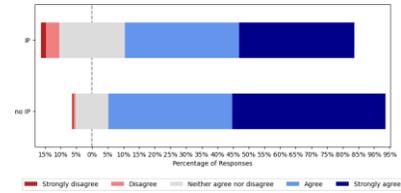

(b) 7. My team supports and values my communication (p-value = 1.14e-05).

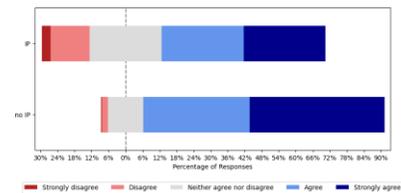

(c) 8. I play an important role in my team (p-value=5.87e-13).

Figure 10: Perceived communication and collaboration.

higher proportion of women software engineers suffer from impostor feelings (60.64%) compared to men (48.82%). The results for races/ethnicities (RQ1.2) indicated higher frequencies of IP in respondents predominantly identifying as Asian (67.85%) and Black or African American (65.11%) software engineers when compared to those identifying as White (50.00%). In the case of underrepresented groups, it is essential to take into account that the work environment may play a role in either triggering or exacerbating these symptoms. For instance, organizations that value psychological safety tend to nurture an environment where people feel encouraged to share ideas without fear of personal judgment.

With respect to IP across various professional roles (RQ1.2), it appears that technical roles, particularly those related to data science, exhibited slightly higher frequencies of impostor feelings. In fact, the data scientist profile, where domain expertise and data-driven insights are pivotal, is typically expected to excel in math/statistics, computing, and business-related skills [15], which seem hard to combine in a single person.

Additionally, we noted that individuals who are married and have children tend to experience IP less frequently than those who are single. These observations suggest potential relationships between personal life circumstances and the prevalence of impostor feelings, which require further investigation.

**RQ2 - Does IP affect the perceived productivity of software engineers?** A consistent pattern emerged for this research question, indicating lower perceived productivity across all five assessed productivity dimensions (RQ2.1 to RQ2.5) of the SPACE developer productivity framework for software engineers suffering from IP. Notably, these differences were statistically significant. These findings provide evidence confirming the hypothesized notion that IP can be a significant barrier to professional productivity.

Recognizing the prevalence and impact of IP as a potential hindrance to productivity, software engineering organizations may consider implementing strategies and support mechanisms to help professionals cope with and overcome these feelings. Such initiatives could include mentorship programs, peer support networks, training in emotional resilience, and fostering a culture of openness and psychological safety.

## 7 THREATS TO VALIDITY

Hereafter, we discuss validity types that are typically considered for survey research [17] and the reliability of our research, as well as mitigation actions taken to address threats and improve the chances

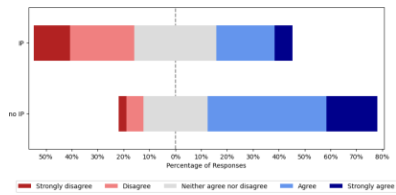

**(a) 9. I am able to focus on and make progress on my work without internal interruptions (p-value=5.75e-23).**

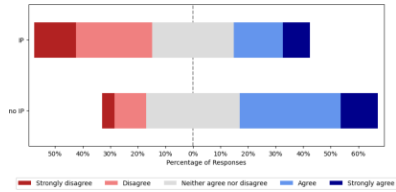

**(b) 10. I am able to focus on and make progress on my work without external interruptions (p-value=1.54e-12).**

**Figure 11: Efficiency and flow.**

of safely concluding that the proposed survey accurately measures what it is supposed to measure.

**Face Validity**. A threat to face validity concerns the unsuitability of the survey instrument for the target audience. To mitigate this threat, we conducted a pilot study to evaluate the instrument, after which we made some minor improvements to the instrument to improve clarity (*e.g.*, adjust numbering and some answer options).

**Content Validity**. To improve content validity, a group of subject matter experts, researchers from universities of three different countries, active with research related to human aspects in software engineering, reviewed and evaluated the questionnaire. One of these experts also had extensive knowledge of the SPACE framework.

**Criterion Validity**. A threat related to criterion validity would be receiving answers from other professionals and not only from software engineers. The survey was sent to representatives of our target population in closed invitation format [24]. Also, we explicitly capture the role of the respondents in the survey, allowing segregated analyses.

**Construct Validity**. The main threat regarding construct validity is using improper instruments. Besides demographics, the survey had two sections, one on IP and one on perceived productivity. For measuring the IP, we made authorized use of the CIPS scale [5], which is the most used scale and has been reported as a reliable instrument[18] and has been used in our related work (*e.g.*, [22, 26]). For perceived productivity, there was no scale for measuring the SPACE productivity framework constructs. Therefore, we designed statements and validated them with subject matter experts.

**Reliability**. With respect to reliability, the main threats would be a potential lack of statistical conclusion validity because of the sample size and the lack of representativeness of the sample. According to [24], for most intents and purposes related to software developers, with a sample size of more than 400, it is possible to claim strong generalizability as long as the representativeness of the sample has also been checked. Our sample of 624 exceeds this suggested sample size. Furthermore, we checked the representativeness by comparing it against the sample from the Stack Overflow Annual Developer survey, for which we observed that the distributions closely resemble, reinforcing our confidence in the quality of the sample. One might argue that the distribution of the countries of the respondents is not representative. This was mainly due to the obligation to use a closed invitation format in order to be allowed to use the CIPS scale. Still, regardless of the country, the developer profile resembles the one from the Stack Overflow survey. Nevertheless, our results and call for actions to prevent IP in software engineers are rather conservative, considering that if we remove the two countries with the most responses (Brazil and Italy), the IP frequency goes up even higher to 65.27% (94 out of 144). Due to space constraints, a complete country-based analysis is out of the scope of this paper and points to future work.

As suggested by [16, 24], and similar to other survey-based studies [23], we used bootstrapping to provide confidence intervals, allowing a more robust and accurate analysis. Our sample and all our analysis procedures are available in our online open science repository[3] and are auditable.

## 8 CONCLUSION

This research represents the first investigation of the prevalence and manifestation of IP in software engineers. We carefully planned and conducted a survey using a validated scale for measuring IP, gathering responses from 624 software engineering professionals. The observed prevalence of IP among software engineers, with 52.72% (CI [48.72, 56.57]) of respondents experiencing frequent to intense levels of IP, highlights the need for increased awareness and support within the software engineering community, especially given that IP may lead to other severe mental disorders [5].

Furthermore, we observed disparities across gender, race/ethnicity, and professional roles. For instance, underrepresented groups, including women and black people, more frequently suffer from IP than men and white people. It is noteworthy that, in the case of underrepresented groups, the work environment may play a role in either inducing or even worsening these symptoms.

Furthermore, we provide statistically significant evidence that software engineers who suffer from IP perceive themselves as less productive than others in terms of their satisfaction and well-being, performance, activity, communication and collaboration, and efficiency and flow. Based on our results, we put forward that software engineering organizations should consider implementing strategies and support mechanisms to promote psychological safety, especially considering underrepresented groups, and help professionals cope with IP feelings and overcome them, ultimately fostering a more inclusive and productive workforce.

## ACKNOWLEDGMENTS

We want to thank each of the 624 software engineers who participated in our survey; without these answers, this research would not have been possible. Financial support was omitted due to the double anonymous review; it will be added later.

# REFERENCES

[1] Deepika Badampudi. 2017. Reporting ethics considerations in software engineering publications. In *2017 ACM/IEEE International Symposium on Empirical Software Engineering and Measurement (ESEM)*. IEEE, 205–210.

[2] Victor R Basili and H Dieter Rombach. 1988. The TAME project: Towards improvement-oriented software environments. *IEEE Transactions on Software Engineering* 14, 6 (1988), 758–773.

[3] Christian Bird, Denae Ford, Thomas Zimmermann, Nicole Forsgren, Eirini Kalliamvakou, Travis Lowdermilk, and Idan Gazit. 2023. Taking Flight with Copilot. *Commun. ACM* 66, 6 (may 2023), 56–62.

[4] Kelly A Cawcutt, Pauline Clance, and Shikha Jain. 2021. Bias, burnout, and imposter phenomenon: the negative impact of under-recognized Intersectionality. *Women's Health Reports* 2, 1 (2021), 643–647.

[5] Pauline Rose Clance. 1985. *The impostor phenomenon: Overcoming the fear that haunts your success*. Peachtree Pub Limited.

[6] Pauline Rose Clance and Suzanne Ament Imes. 1978. The imposter phenomenon in high achieving women: Dynamics and therapeutic intervention. *Psychotherapy: Theory, research & practice* 15, 3 (1978), 241.

[7] Lindsay Duncan, Gita Taasoobshirazi, Ashana Vaudreuil, Jitendra Sai Kota, and Sweta Sneha. 2023. An evaluation of impostor phenomenon in data science students. *International journal of environmental research and public health* 20, 5 (2023), 4115.

[8] Denae Ford and Chris Parnin. 2015. Exploring causes of frustration for software developers. In *2015 IEEE/ACM 8th International Workshop on Cooperative and Human Aspects of Software Engineering*. IEEE, 115–116.

[9] Nicole Forsgren, Margaret-Anne Storey, Chandra Maddila, Thomas Zimmermann, Brian Houck, and Jenna Butler. 2021. The SPACE of developer productivity. *Commun. ACM* 64, 6 (2021), 46–53.

[10] Nicole Forsgren, Margaret-Anne Storey, Chandra Maddila, Thomas Zimmermann, Brian Houck, and Jenna Butler. 2021. The SPACE of Developer Productivity: There's more to it than you think. *ACM Queue* 19, 1 (2021), 20–48.

[11] Daniela Girardi, Nicole Novielli, Davide Fucci, and Filippo Lanubile. 2020. Recognizing developers' emotions while programming. In *Proceedings of the ACM/IEEE 42nd International Conference on Software Engineering*. 666–677.

[12] Daniel Graziotin, Fabian Fagerholm, Xiaofeng Wang, and Pekka Abrahamsson. 2017. Consequences of unhappiness while developing software. In *2017 IEEE/ACM 2nd International Workshop on Emotion Awareness in Software Engineering (SEmotion)*. IEEE, 42–47.

[13] Daniel Graziotin, Fabian Fagerholm, Xiaofeng Wang, and Pekka Abrahamsson. 2018. What happens when software developers are (un) happy. *Journal of Systems and Software* 140 (2018), 32–47.

[14] Watts S Humphrey. 1988. The software engineering process: definition and scope. In *Proceedings of the 4th International Software Process Workshop on Representing and Enacting the Software Process*. 82–83.

[15] Miryung Kim, Thomas Zimmermann, Robert DeLine, and Andrew Begel. 2017. Data scientists in software teams: State of the art and challenges. *IEEE Transactions on Software Engineering* 44, 11 (2017), 1024–1038.

[16] Skylar Lei and Michael R Smith. 2003. Evaluation of several nonparametric bootstrap methods to estimate confidence intervals for software metrics. *IEEE Transactions on Software Engineering* 29, 11 (2003), 996–1004.

[17] Johan Linaker, Sardar Muhammad Sulaman, Martin Höst, and Rafael Maiani de Mello. 2015. Guidelines for conducting surveys in software engineering. *Lund University* 50 (2015).

[18] Karina KL Mak, Sabina Kleitman, and Maree J Abbott. 2019. Impostor phenomenon measurement scales: a systematic review. *Frontiers in Psychology* 10 (2019), 671.

[19] A. Mani and Rebeka Mukherjee. 2016. A study of FOSS 2013 survey data using clustering techniques. In *2016 IEEE International WIE Conference on Electrical and Computer Engineering (WIECON-ECE)*. 118–121. https://doi.org/10.1109/WIECON-ECE.2016.8009099

[20] Stack Overflow. 2023. *Stack Overflow Annual Developer Survey*. Retrieved September 8, 2023 from https://insights.stackoverflow.com/survey

[21] Gregorio Robles, Laura Arjona Reina, Jesús M González-Barahona, and Santiago Dueñas Domínguez. 2016. Women in free/libre/open source software: The situation in the 2010s. In *Open Source Systems: Integrating Communities: 12th IFIP WG 2.13 International Conference, OSS 2016, Gothenburg, Sweden, May 30-June 2, 2016, Proceedings 12*. Springer, 163–173.

[22] Adam Rosenstein, Aishma Raghu, and Leo Porter. 2020. Identifying the prevalence of the impostor phenomenon among computer science students. In *Proceedings of the 51st ACM Technical Symposium on Computer Science Education*. 30–36.

[23] Stefan Wagner, Daniel Méndez Fernández, Michael Felderer, Antonio Vetrò, Marcos Kalinowski, Roel Wieringa, Dietmar Pfahl, Tayana Conte, Marie-Therese Christiansson, Desmond Greer, et al. 2019. Status quo in requirements engineering: A theory and a global family of surveys. *ACM Transactions on Software Engineering and Methodology (TOSEM)* 28, 2 (2019), 1–48.

*[24]* Stefan Wagner, Daniel Mendez, Michael Felderer, Daniel Graziotin, and Marcos Kalinowski. 2020. Challenges in survey research. *Contemporary Empirical Methods in Software Engineering* (2020), 93–125.

[25] Taro Yamane. 1973. *Statistics. An introductory analysis. Third edition*. Harper & Row.

[26] Angela Zavaleta Bernuy, Anna Ly, Brian Harrington, Michael Liut, Andrew Petersen, Sadia Sharmin, and Lisa Zhang. 2022. Additional Evidence for the Prevalence of the Impostor Phenomenon in Computing. In *Proceedings of the 53rd ACM Technical Symposium on Computer Science Education V. 1*. 654–660.